\documentclass[aip,apl,preprint]{revtex4-2}

\usepackage{amsmath}
\usepackage{graphicx}	
\usepackage{hyperref}
\usepackage[T1]{fontenc}
\usepackage{textcomp}
\usepackage{bm}	
\usepackage{color}

\renewcommand{\deg}{$^\circ$}

\begin{document}

\title{Compact tunable YIG-based RF resonators}

\author{Jos\'e Diogo Costa}
\email[Author to whom correspondence should be addressed. E-mail: ]{diogo.costa@imec.be}
\author{Bruno Figeys}
\author{Xiao Sun}
\author{Nele Van Hoovels}
\author{Harrie A. C. Tilmans}
\author{Florin Ciubotaru}
\author{Christoph Adelmann}
\email[]{christoph.adelmann@imec.be}
\affiliation{Imec, 3001 Leuven, Belgium}

\begin{abstract}
We report on the design, fabrication, and characterization of compact tunable yttrium iron garnet (YIG) based RF resonators based on {\textmu}m-sized spin-wave cavities. Inductive antennas with both ladder and meander configurations were used as transducers between spin waves and RF signals. The excitation of ferromagnetic resonance and standing spin waves in the YIG cavities led to sharp resonances with quality factors up to 350. The observed spectra were in excellent agreement with a model based on the spin-wave dispersion relations in YIG, showing a high magnetic field tunability of about 29 MHz/mT. 
\end{abstract}

\maketitle

Yttrium iron garnet (YIG, Y$_3$Fe$_5$O$_{12}$) radiofrequency (RF) resonators based on YIG spheres are a well-established technology and are employed in a broad range of applications, including RF filters and oscillators.\cite{K_1948,D_1958,FB_1959,C_1961,H_1985} These technologies benefit from the low intrinsic magnetic loss of YIG, leading to quality ($Q$) factors  of up to several thousand. Moreover, the resonance frequency of filters and oscillators can be tuned over a wide frequency by means of an applied magnetic field. Yet, YIG spheres have mm-scale diameters and require even larger transducers for input and output RF signals in addition to the system to apply magnetic fields.

More recently, the development of thin film deposition techniques, in particular liquid phase epitaxy (LPE),\cite{HBH_1973} has led to an increased interest in planar YIG resonators.\cite{IC_1986,I_1988} This has resulted in considerable reduction of size and form factor,\cite{GW_1975,V_1980,APO_1978,WSC_1977} culminating in $Q$ factors reaching several thousand for groove-based cavities.\cite{CAB_1977,OSS_1978,CH_1985} However, the area of such devices was still in the square mm range, with YIG thicknesses of several {\textmu}m, and usually operating in a flip-chip configuration. 

In the last decade, the field of magnonics has seen tremendous progress and spin-wave devices have been miniaturized to {\textmu}m and sub-{\textmu}m dimensions.\cite{SCH_2010,DS_2013,CVS_2015,SVM_2017,MCV_2020,TDT_2020} Nonetheless, while YIG micro- and nanostructures have been studied intensively in recent years,\cite{HVL_2014,ZCF_2017,HBS_2020,WKS_2020} the miniaturization of YIG-based filters and resonators to sub-mm dimensions has received little attention so far.\cite{YWL_2013, DBW_2020} Here, we report on the fabrication and characterization of YIG thin film RF resonators that are based on spin-wave cavities with {\textmu}m lateral dimensions. This approach allows for a compact device design, with $Q$ factors of up to 350 and large magnetic field tunability. A model of the spin-wave dispersion relations of the structures is in excellent agreement with the observed resonator spectra. These results show such devices are promising to reduce the footprint of YIG-based filters, while maintaining high $Q$ factors and tunable electrical output.

The devices were based on 800 nm thick (111) YIG films (Innovent Technologieentwicklung, Jena, Germany) deposited by LPE on (111) gadolinium gallium garnet (GGG) substrates. The films showed low Gilbert damping of $\alpha \approx 1 \times 10^{-4}$.\cite{DSL_2017} A combination of e-beam lithography, wet etching, and lift-off was used for device fabrication. The spin-wave cavities were microfabricated by wet etching (H$_3$PO$_4$, 130\deg C) using a SiO$_2$ hardmask. After planarization by spin-on carbon, the 100 nm thick Au antenna transducers were defined by a lift-off process.

\begin{figure}[t]
\begin{center}
	\includegraphics[width=12cm]{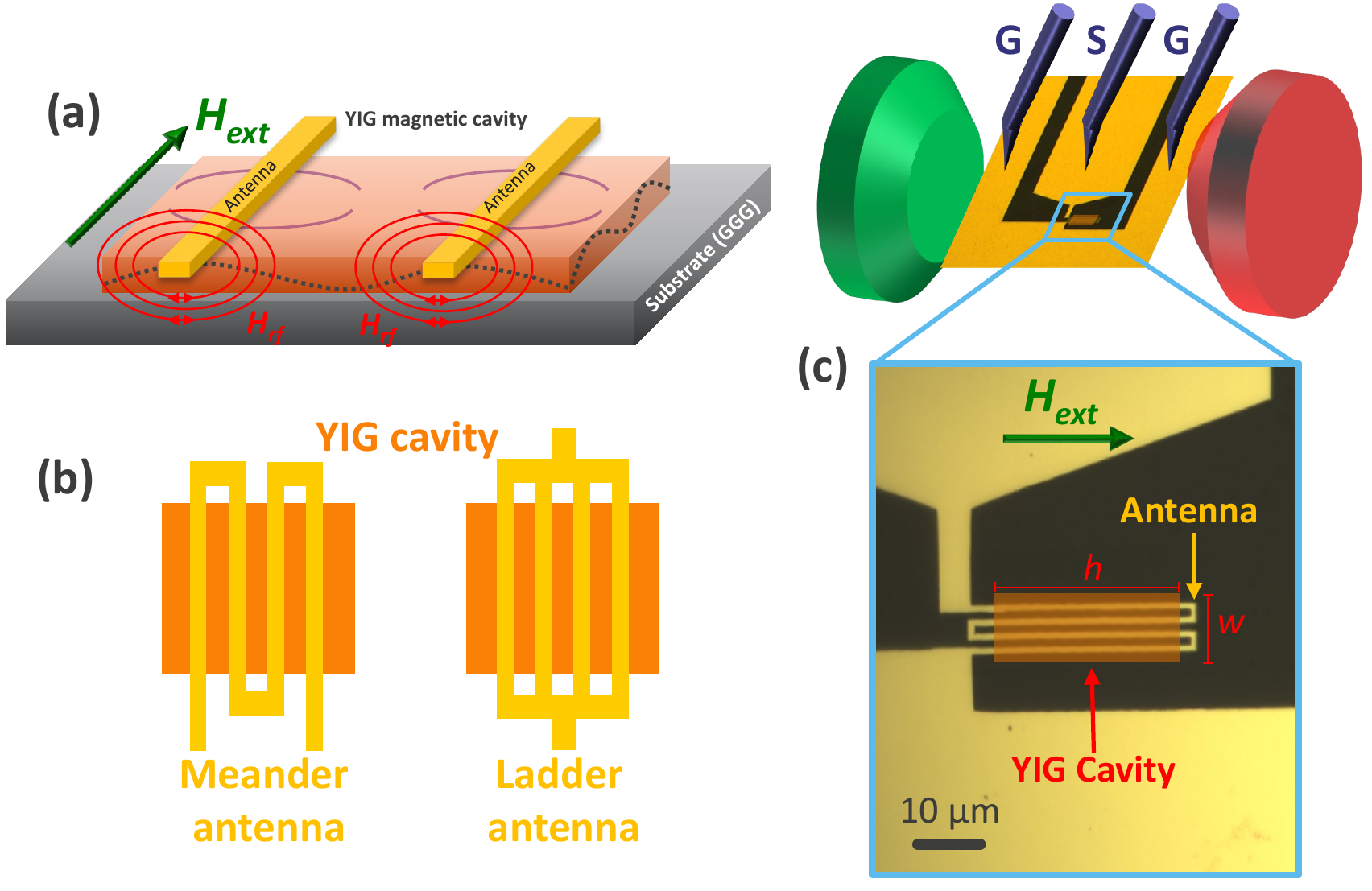}
	\end{center}
	\caption {(a) Schematic of a YIG cavity resonator. Examples of standing spin-wave modes inside the YIG cavity are represented by dashed lines. (b) Inductive antennas used in this study: meander and ladder antennas. (c) Schematic of experimental setup (top), including the probing scheme and the external electromagnet, as well as optical micrograph (bottom) of a meander antenna resonator. The red box indicates the position of the YIG cavity with width $w$ and height $h$.}
	\label{fig:schematic}
\end{figure}

Figure \ref{fig:schematic}(a) depicts a schematic of a YIG cavity resonator along with the direction of the applied external magnetic field. Spin-wave cavity modes (dashed lines) are excited inside the cavity by inductive antennas, which generate oscillating Oersted fields from RF currents. Two distinct types of antennas were used in this study: meander and ladder structures [Fig.~\ref{fig:schematic}(b)]. Both meander and ladder antennas consisted each of $N$ identical wires [$N = 4$ in Fig.~\ref{fig:schematic}(b)] with a width of 1 {\textmu}m and a pitch of 2 {\textmu}m. Figure \ref{fig:schematic}(c) shows an optical micrograph of processed device including a meander antenna. The dashed red line indicates the region where the YIG cavity ($8 \times 32$ {\textmu}m$^2$) is located. The difference between the two types of antennas lies in the direction of the RF current: while in ladder antennas, currents in wires flow in the same direction with the same phase, currents in adjacent meander antenna wires have a phase shift of $\pi$, \emph{i.e.} they flow in opposite directions. Hence, local magnetic fields are always in the same direction for ladder antennas but alternating for meander antennas. 

\begin{figure}[t]
\begin{center}
	\includegraphics[width=14cm]{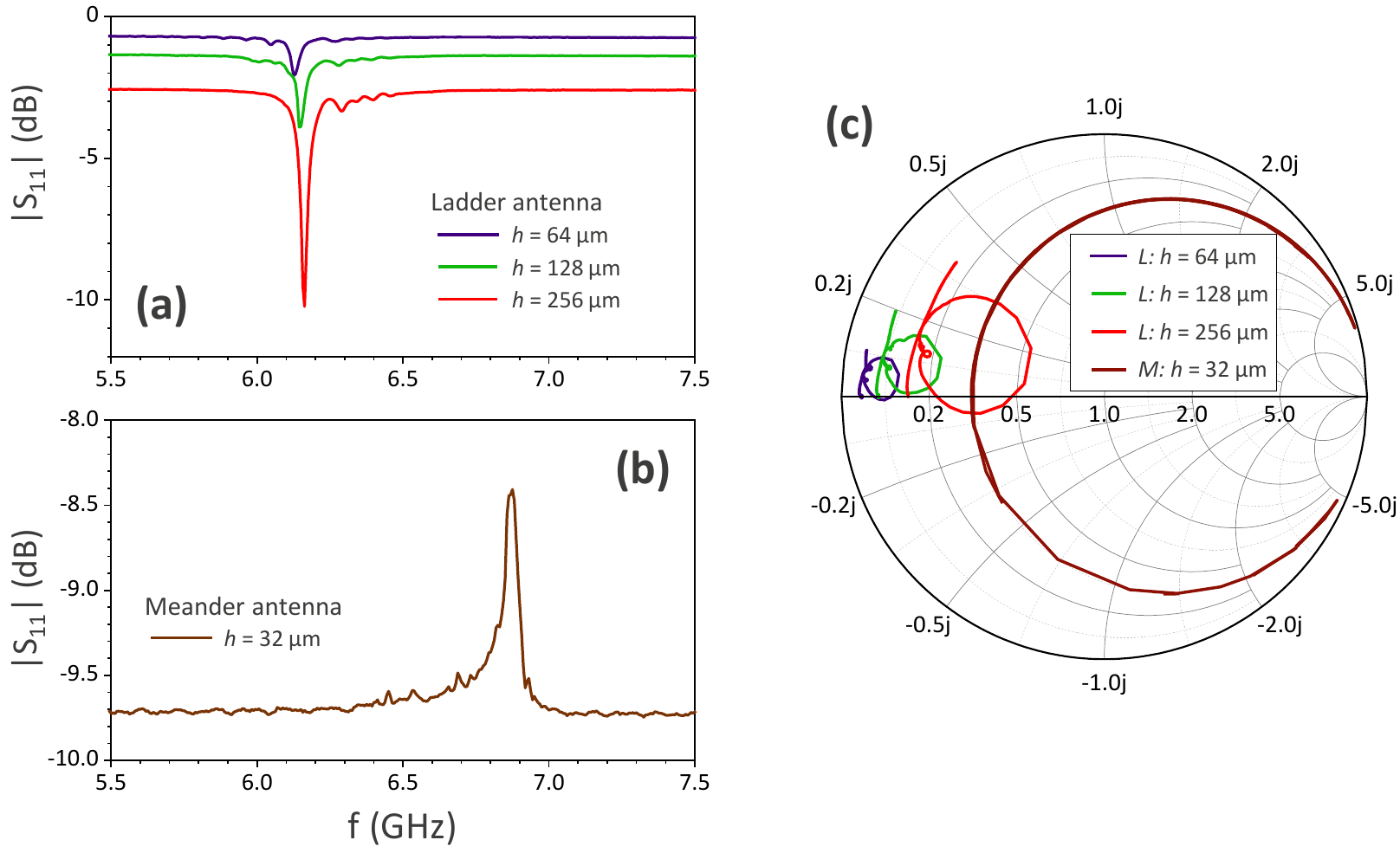}
	\end{center}
	\caption{RF characteristics of YIG cavity resonators. Measured magnitude of the $S_{11}$-parameter \emph{vs.} frequency  for resonators with (a) ladder antennas (L) and (b) a meander antenna (M), for indicated cavity heights $h$ ($w = 16$ {\textmu}m). (c) Smith chart representation of the RF measurements matched to a 50 $\Omega$ impedance ($f = 10$ MHz to 15 GHz). For ladder antennas, external bias field $\mu_0 H_\mathrm{ext} = 145$ mT; for the meander antenna, $\mu_0 H_\mathrm{ext} = 163$ mT.}
	\label{Fig:Smith}
\end{figure}

The RF response of the devices was assessed by measuring the RF reflection ($S_{11}$-parameter) with a Keysight E8363B network analyzer (input RF power $-17$ dBm) at frequencies between 10 MHz and 15 GHz. The antennas were connected between signal and ground paths of 50 $\Omega$ coplanar waveguides and contacted by GSG probes [see Fig.~\ref{fig:schematic}(c)]. During the measurements, an external magnetic field was applied along the antenna wires [see Fig.~\ref{fig:schematic}(c)] using an electromagnet, calibrated by a Hall effect Gauss meter. Figures \ref{Fig:Smith}(a) and (b) show as examples the return loss of three ladder-antenna resonators ($N = 8$) with different cavity heights ($w = 16$ {\textmu}m) as well as of a meander-antenna resonator ($N = 8$, $w = 16$ {\textmu}m, $h = 32$ {\textmu}m), respectively, after de-embedding the contact pad parasitics. Sharp resonances are observed at frequencies between 6.1 and 6.2 GHz, next to a series of weaker resonances. A Smith chart representation of the device impedances is shown in Fig.~\ref{Fig:Smith}(c). The intrinsic $Q$ factor of the ladder-antenna devices varied between about 200 and 350, which was obtained for the largest device. The RF absorption by the YIG cavity increased with $h$ due to the increasing magnetic volume. By contrast, the $Q$ factor for the meander-antenna resonator was lower, about 120.

\begin{figure*}[t]
\begin{center}
	\includegraphics[width=15.5cm]{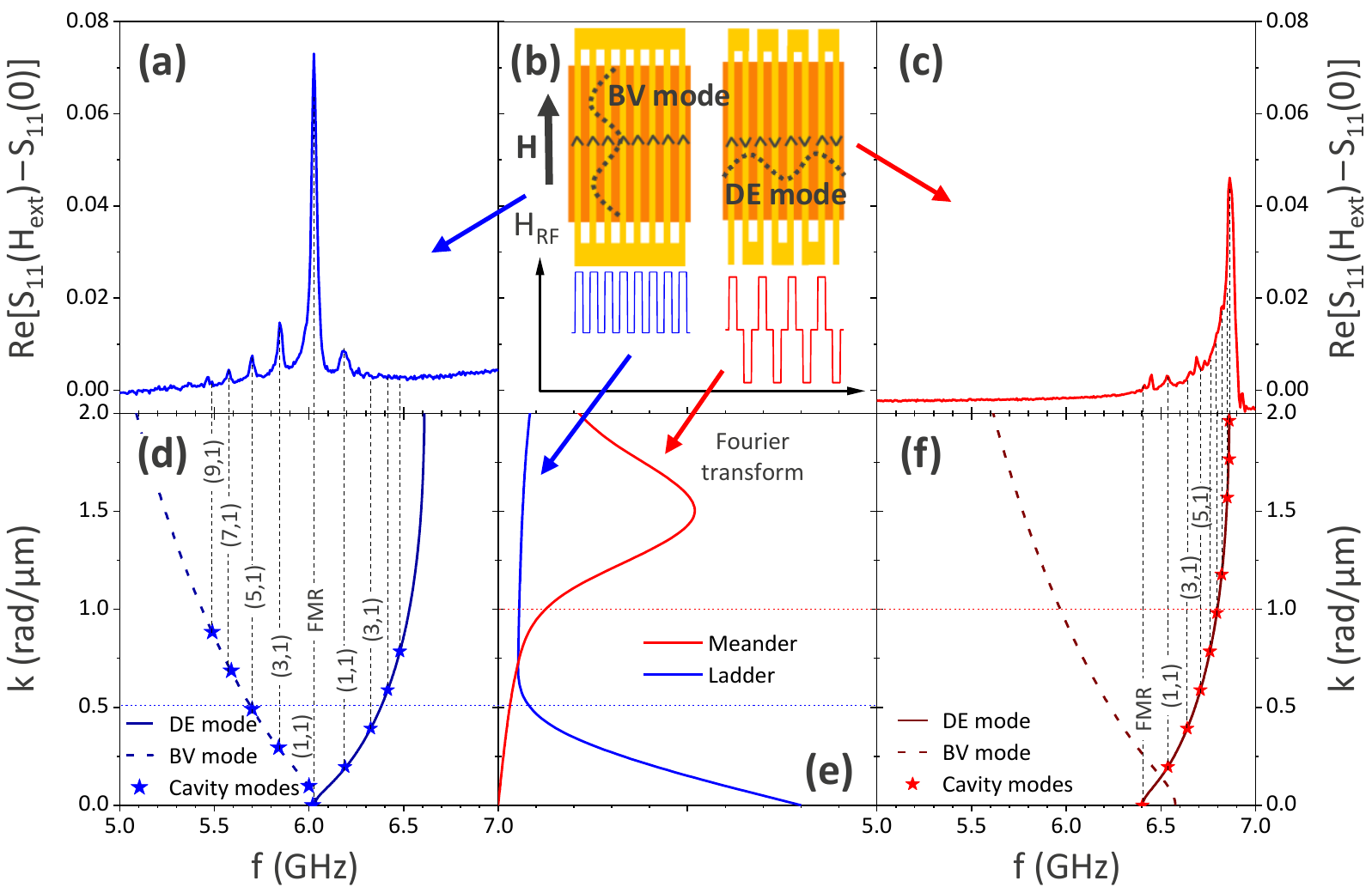}
	\end{center}
	\caption {RF characteristics of YIG cavity resonators ($16 \times 32$ {\textmu}m$^2$ cavity size) with (a) ladder and (c) meander antennas. (b) schematic representation of the studied devices (top), the exciting Oersted field (bottom), as well as BV and DE spin-wave configurations. Dispersion relations for the resonators with (d) ladder and (f) meander antennas, with DE (solid lines, $m = 1$), BV (dashed lines, $m = 1$), and excited cavity modes (stars) with mode numbers $(n,m)$. $\mu_0 H_\mathrm{ext} = 145$ and 163 mT for ladder and meander antenna resonators, respectively. (e) Excitation spectrum due to the Oersted field for ladder (blue line) and meander antennas (red line).}
	\label{fig:spectra1}
\end{figure*}

We now discuss in more detail the different resonance spectra for the devices including both ladder and meander antennas. Resonator spectra are shown in Figs.~\ref{fig:spectra1}(a) and \ref{fig:spectra1}(c) for ladder ($\mu_0 H_\mathrm{ext} = 145$ mT) and meander ($\mu_0 H_\mathrm{ext} = $163 mT) antennas, respectively. The cavity area was $w\times h = 16 \times 32$ {\textmu}m$^2$ in both cases and all applied magnetic fields were sufficient to saturate the magnetization. In addition to the main resonance, the spectra showed several additional weaker resonances that are well separated in the case of the ladder antenna. This behavior can be understood by considering the spin-wave dispersion relations in a planar rectangular YIG cavity with dimensions $\ell_1\times \ell_2$, which are given by\cite{MCV_2020,KS_1986} 

\begin{equation}
\label{eq:Disp_Rel}
f_\mathrm{n,m} = \frac{1}{2\pi}\sqrt{(\omega_0 + \omega_\mathrm{M}\lambda_\mathrm{ex}k_\mathrm{tot}^2) (\omega_0+\omega_\mathrm{M}\lambda_\mathrm{ex} k_\mathrm{tot}^2 + \omega_M F)}\,,
\end{equation}
\noindent with $\omega_0=\gamma \mu_0 H_\mathrm{ext}$, $\omega_\mathrm{M}=\gamma \mu_0 M_s$, and the abbreviations
\begin{equation}
\label{eq:F}
F = P + \sin^2\phi\times \bigg( 1-P(1+\cos^2(\theta_\mathrm{k} - \theta_\mathrm{M})) + \left. \frac{\omega_\mathrm{M} P(1-P)\sin^2(\theta_\mathrm{k} - \theta_\mathrm{M})}{\omega_0 + \omega_\mathrm{M}\lambda_\mathrm{ex} k_\mathrm{tot}^2} \right)  \,,
\end{equation}
\begin{equation}
\label{eq:g}
P = 1- \frac{1-e^{-dk_\mathrm{tot}}}{dk_\mathrm{tot}}\,.
\end{equation} 
\noindent Here, $k_\mathrm{tot} = \sqrt{k_\mathrm{n}^2 + k_\mathrm{m}^2}$ with $k_\mathrm{n} =  n\pi/\ell_1$ and $k_\mathrm{m} =  m\pi/\ell_2$ the quantized wavenumbers in the two cavity directions. $n$ and $m$ are the mode numbers. $\phi$ denotes the angle between the magnetization and the normal to the waveguide, $\theta_\mathrm{M}$ is the angle between the magnetization and the direction of $k_\mathrm{n}$, and $\theta_\mathrm{k} = \arctan(k_\mathrm{m}/k_\mathrm{n})$. 

In our devices, the symmetry between the two confinement directions is broken by the direction of the magnetic field and two types of spin-wave modes need to be distinguished: (i) confined backward volume (BV) modes with the direction of $k_\mathrm{n}$ parallel to $H_\mathrm{ext}$, $\ell_1=h$, $\ell_2=w$, $\phi = \pi/2$, $\theta_\mathrm{M} = 0$; and (ii) confined Damon-Eshbach (DE) modes with the direction of $k_\mathrm{n}$ perpendicular to $H_\mathrm{ext}$, $\ell_1=w$, $\ell_2=h$, $\phi = \pi/2$, $\theta_\mathrm{M} = \pi /2$ [see Fig.~\ref{fig:spectra1}(b)]. Frequencies and wavenumbers of resulting discrete modes for different modes with $(n,m = 1)$ (saturation magnetization $M_s = 130$ kA/m, exchange constant $A = 3.5$ pJ/m, $\alpha = 1\times 10^{-4}$) are represented in Figs.~\ref{fig:spectra1}(d) and \ref{fig:spectra1}(f) for the two devices and $\mu_0 H_\mathrm{ext} = 145$ mT and 163 mT, respectively. In addition, Figs.~\ref{fig:spectra1}(d) and \ref{fig:spectra1}(f) also show the continuous dispersion relations of BV and DE spin waves (blue/red dashed and solid lines, respectively), from which the cavity modes are derived. 

A comparison with the experimental spectra in Figs. \ref{fig:spectra1}(a) and (c) shows excellent agreement with the frequencies of both BV and DE spin-wave cavity modes with $m = 1$. Further insight into the excited resonances, their relative amplitudes, and the dependence on the antenna design can be gained by considering the excitation efficiency of a spin-wave (cavity) mode by the Oersted field from an inductive antenna, which is given by\cite{DD_2015,MCV_2020}
\begin{equation}
\label{eq:eff_overlap}
\Gamma_n \propto \left| \int_V \bm{H}_{\mathrm{RF}}(\bm{x}) \cdot \bm{m}(\bm{x}) \: d^3x \right|,
\end{equation}
\noindent with $\bm{H}_{\mathrm{RF}}(\bm{x})$ the exciting Oersted field, $\bm{m}(\bm{x})$ the dynamic magnetization of the spin-wave mode, and $V$ the cavity volume. In wavevector space, the excitation efficiency is thus given by a Fourier transform of the magnetic excitation field. To a first approximation, the magnetic field underneath a wire antenna with width $d$ can be written as $H_{\mathrm{RF}} \approx \pm I_{RF}/2d$ in the wire region and 0 outside, as illustrated in Fig.~\ref{fig:spectra1}(b) for both ladder and meander antennas. Here, $I_{RF}$ is the RF current. 

The resulting spatial Fourier spectra of the excitation fields transverse to the wires are shown in Fig.~\ref{fig:spectra1}(e) for both ladder and meander antennas ($N = 8$). This direction corresponds to $\theta_\mathrm{M} = \pi /2$ and thus to the excitation of DE-like spin-wave cavity modes (see above). The spectra show that ladder antennas efficiently excite DE cavity modes with small mode numbers (small wavenumbers $k$, large wavelengths) but cannot excite modes with higher $k$. A comparable result is found for the Fourier transform along the wires of the ladder antenna (not shown), which describes the coupling to BV cavity modes. As a result, the main resonance in the experimental spectrum in Fig.~\ref{fig:spectra1}(a) can be attributed to a superposition of the DE and BV ferromagnetic resonances (FMR)---which are nondegenerate due to the finite dimensions of the rectangular YIG cavity---and the first BV spin-wave cavity mode $(n = 1,m = 1)$. Additional resonances at lower frequency correspond to higher order BV spin-wave modes with increasing $n$, whereas only one higher-order DE mode was clearly observed at higher frequencies. We note that due to symmetry reasons, only odd BV and DE cavity modes can be excited. BV modes with $m = 3$ follow a dispersion relation that is approximately 200 MHz above the $m = 1$ curve. Therefore, the broadening of the peak at $\sim 6.2$ GHz may be attributed to the superposition with the $(n = 1,m = 3)$ mode.  Modes with $m > 1$ and $n > 1$ cannot be distinguished as they are superposed to the main resonance as well as the $m = 1$ modes, and their intensity decreases rapidly with increasing $n$ (as for $m = 1$) and $m$.

By contrast, meander antennas preferentially excite DE cavity modes with larger wavenumbers around a maximum determined by the wire pitch. Moreover, due to the opposite directions of the magnetic fields underneath adjacent wires, the meander antenna cannot excite BV modes since the average magnetic field transverse to the wires is zero. Therefore, the spectrum consists of DE modes with increasing mode number $n$ until the dispersion relation becomes flat at high wavenumbers. The main resonance in the spectrum in Fig.~\ref{fig:spectra1}(c) thus consists of a superposition of a large number of BV cavity modes large $n$ and thus nearly continuous $k$. In addition, modes with large (odd) $m$ can also be expected to contribute to the main resonance. As a result, the $Q$ factor of the main resonance is lower for meander antennas than for ladder antennas. Note that the $Q$ factor of resonators with meander antennas can be optimized by reducing the wire pitch, which reduces the excitation of DE cavity modes with low wavenumbers.

\begin{figure}[t]
\begin{center}
	\includegraphics[width=8.5cm]{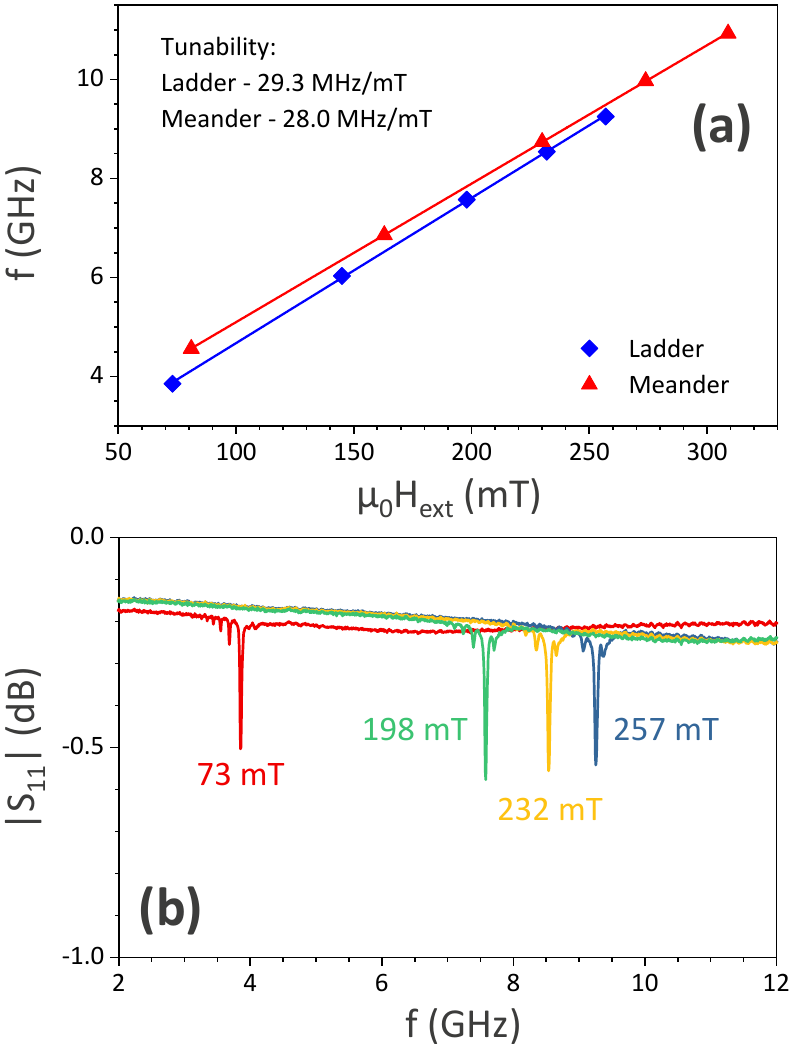}
	\end{center}
	\caption {Tunability of YIG cavity resonators. (a) Resonance frequency \emph{vs.} external magnetic bias field for resonators ($16 \times 32$ {\textmu}m$^2$ cavity) with ladder and meander antennas, as indicated. The solid lines represent best linear fits to the data. (b) Magnitude of the experimental $S_{11}$-parameter \emph{vs.} frequency for the ladder antenna resonator for different magnetic bias fields.}
	\label{fig:tunability}
\end{figure}

One of the key advantages of YIG resonators is their tunability by a magnetic field. This is illustrated in Fig.~\ref{fig:tunability}(a), which shows the dependence of the measured main resonance frequency on the applied magnetic field for ladder and meander antennas. In both cases, the dependence on the studied magnetic field range was linear in the studied range with slopes of 29.3 MHz/mT (ladder) and 28.0 MHz/mT (meander). The tunability was very similar to that of devices based on bulk YIG,\cite{H_1985,CI_1986} demonstrating the device miniaturization does not affect the tunability. The slight dependence on the antenna design can be attributed to the different magnetic field dependence of the relevant spin-wave cavity modes. Figure \ref{fig:tunability}(b) shows the corresponding measured return loss for the ladder-antenna resonator, indicating that the device impedance varies only weakly with external bias field.

\begin{figure}[t]
\begin{center}
	\includegraphics[width=8.5cm]{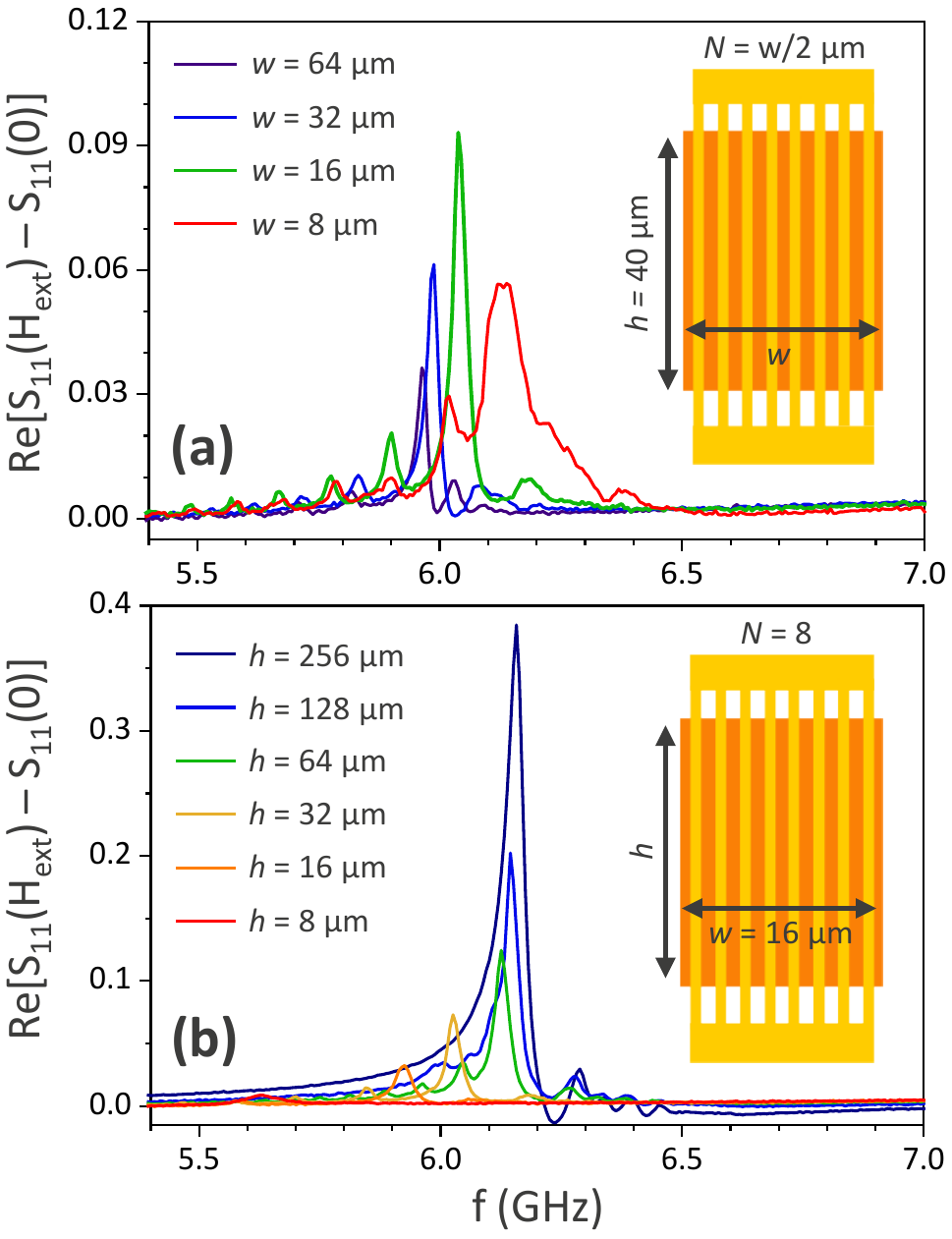}
	\end{center}
	\caption{RF characteristics of YIG cavity resonators with ladder antennas for different cavity dimensions. (a) Different cavity widths $w$ for constant antenna wire density ($N = w/2$ {\textmu}m) and cavity height ($h = 40$ {\textmu}m). (b) Different cavity heights for constant cavity width ($w = 16$  {\textmu}m, $N = 8$). In all cases, $\mu_0 H_\mathrm{ext} = 145$ mT.}
	\label{fig:5}
\end{figure}

These results indicate that ladder antennas lead to a better-defined device response with a sequence of well-marked and sharp resonances. In the following, we focus on the signal optimization of ladder structures. Figure~\ref{fig:5}(a) shows the frequency response of resonators with identical antenna wire width (1 {\textmu}m) and pitch (2 {\textmu}m), identical cavity height $h = 40$ {\textmu}m, but different cavity width $w$. Maintaining a constant wire density for increasing $w$ was achieved by setting the number of wires in each resonator to $N = w/2$ {\textmu}m. In this case, increasing $w$ leads to two competing effects: (i) an increase of the magnetic volume and transducer size; and (ii) the redistribution of the total current in an increasing number of parallel wires, which lowers the Oersted field underneath each individual wire. Whereas (i) increases the RF absorption by the cavity, (ii) reduces the external excitation. Both effects tend to compete with each other and, as a result, an optimum width of $w = 16$ {\textmu}m ($N = 8$) was observed, as shown in Fig.~\ref{fig:5}(a). As expected, the separation between adjacent DE cavity modes decreased for larger $w$ and several peaks became superimposed for the largest cavity. Narrower cavities also showed reduced resonator $Q$ factors, possibly due to edge effects or processing imperfections.

The effect of varying the cavity height $h$ is illustrated in Fig.~\ref{fig:5}(b). In this case, longer antennas overlap with a larger magnetic volume without reducing the driving Oersted field, leading to increased RF absorption by the longer YIG cavity. Thus, the strongest RF absorption was obtained for the largest cavity with $h = 256$ {\textmu}m, in keeping with the results in Fig.~\ref{Fig:Smith}(a). The main resonance frequency increased slightly with $h$, which may be attributed to small changes in the internal magnetic field resulting in the observed shift of the dispersion relation as a function of the cavity length.\textcolor{red}{\cite{DS_2013}}

In conclusion, we have studied the characteristics of RF resonators based on YIG cavities with areas down to 128 {\textmu}m$^2$ ($\approx 10^{-4}$ mm$^2$). Both ladder- and meander-type antennas were used as transducers between the RF and spin-wave domains. The resonators showed $Q$ factors up to 350, depending on the YIG cavity dimensions, and a magnetic field tunability of about 29 MHz/mT. The reduced Q factor with respect to bulk or mm-size YIG resonators can be attributed to the combination of larger intrinsic damping of thin film YIG as well as edge effects  at the cavity boundaries. The observed frequency dependence of the resonators was in excellent agreement with the spin-wave dispersion relations in YIG, indicating that the different transducers excited distinctly different cavity modes. Concretely, the model indicated that ladder antennas mainly coupled to ferromagnetic resonance, whereas meander antennas excited standing spin-wave modes with large mode numbers $n$. 

A figure of merit of resonators for RF filters can be defined as $\mathrm{FoM} = k^2_\mathrm{eff} \times Q_u$ with $Q_u$ the unloaded $Q$ factor of the resonance and $k^2_\mathrm{eff}$ the effective coupling coefficient, which can be deduced from the admittance $Y_{11}$ of the resonator. Acoustic resonators of similar size used in RF filers possess FoMs of 50--200.\cite{LCZ_2020} By contrast, our resonators have $\mathrm{FoMs} \approx 5$--10, which is still an order of magnitude smaller, mainly due to the intrinsically smaller energy transfer from inductive antennas into the spin-wave system with respect to (piezoelectric) transducers used in acoustic filters. Future improvements of $k^2_\mathrm{eff}$ may come from the use of magnetoelectric transducers, which still remain to be brought to maturity.\cite{MCV_2020} However, to cover a large amount of RF bands on one chip, many different filters typically need to be co-packaged into one system. By contrast, one to several tunable spin-wave resonators integrated in single filter can replace a large filter bank, consisting of many acoustic resonators, and therefore compensate for the lower FoM. The results thus show that {\textmu}m-sized YIG resonators may find applications in future miniaturized magnetically tunable RF filters. The small device size and form factor of the resonators are also of interest for future integrated systems in a package combining different RF components. 

All data needed to support the conclusions are present in the paper. Additional data may be requested from the authors. This work has received funding from the imec.xpand fund. The authors would like to thank Patrick Vandenameele and Peter Vanbekbergen for their support of the project as well as Xavier Rottenberg, Kristof Vaesen, and Barend van Liempd for many valuable discussions. J.D.C. acknowledges financial support from the European Union MSCA-IF Neuromag under grant agreement No. 793346. FC's and CA's contributions have been funded in part by the European Union's Horizon 2020 research and innovation program within the FET-OPEN project CHIRON under grant agreement No. 801055.

\end{document}